\documentclass[10pt,a4paper]{article}

\usepackage{fourier}
\usepackage{graphicx}
\usepackage[latin1]{inputenc}
\usepackage{latexsym}
\usepackage{natbib}
\usepackage{hyperref}

\begin{document}

\title{Uncertainties in the Sunspot Numbers: \\Estimation and Implications}
\author{Thierry Dudok de Wit$^{1)}$ \and Laure Lef\`evre$^{2)}$ \and Fr\'ed\'eric Clette$^{2)}$}
\date{\small  $^{1)}$ LPC2E, CNRS and University of Orl\'eans, 3A Av. de la Recherche Scientifique, 45071 Orl\'eans cedex 2, France \ \ $^{2)}$ Royal Observatory of Belgium, 3 avenue Circulaire, 1180 Brussels, Belgium \\ \bigskip
This is a slightly updated version of an article published in Solar Physics (2016), and available via \url{http://dx.doi.org/10.1007/s11207-016-0970-6}}
\maketitle

\begin{abstract} 
\noindent
Sunspot number series are subject to various uncertainties, which are still poorly known. The need for their better understanding was recently highlighted by the major makeover of the international Sunspot Number \citep{clette14}. We present the first thorough estimation of these uncertainties, which behave as Poisson-like random variables with a multiplicative coefficient that is time- and observatory-dependent. We provide a simple expression for these uncertainties, and reveal how their evolution in time coincides with changes in the observations, and processing of the data. Knowing their value is essential for properly building composites out of multiple observations, and for preserving the stability of the composites in time.
\end{abstract}

\section{Introduction}
\label{sec_introduction}

\subsection{Historical Context}

The sunspot number time series is the longest still-ongoing scientific experiment, and is also our only direct observation of solar activity up to centennial time-scales. As such, it is of major importance for quantifying the influence of the Sun on the heliosphere.

The Wolf number $\mathrm{N_W}$, also known as the sunspot number, or relative sunspot number, was introduced by Rudolf Wolf in 1848 \citep{wolf50}. This quantity is based on the total number of sunspots $\mathrm{N_s}$ and the number $\mathrm{N_g}$ of sunspot groups that are present on the Sun according to the formula (whose different weights are at the origin of the qualifier \textit{relative}):
\begin{equation}
\label{eq_Wolf}
N_{\mathrm{W}} = k(10 N_{\mathrm{g}} + N_{\mathrm{s}}) \ .
\end{equation}
This combination is justified by the fact that neither of the two numbers, by itself, satisfactorily describes solar activity. The scaling factor $\mathrm{k}$ is assigned to each observer to compensate for their differing observational qualities; its value mainly depends on the observer's ability to detect the smallest sunspots (in relation to telescope aperture, local seeing, and personal experience) and on how complex groups are split. The method used today for computing these scaling factors is explained by \citet{clette07}.

% \subsection{The sunspot number: history}

Hundreds of observers have collected and tabulated  Wolf numbers $\mathrm{N_W}$ over various periods in time. Merging  all of these records into one single composite sunspot number $S_{\mathrm{N}}$ is a formidable enterprise \citep{bray64,hoyt98,clette07}, which requires a good understanding of their uncertainties. Surprisingly, very little is known about these uncertainties. So far, they have never been included in the original Wolf number records. Our objective is to fill that gap by providing a better understanding of these uncertainties, and a way to quantify them. 

Before addressing them, however, let us first address how observations and conventions have evolved over time.

Routine sunspot observations started in the early 17th century by astronomers such as Harriot, Galileo, and Fabricius. Thanks to them, and to multiple other observers, it is now possible to reconstruct an index of solar activity over more than four centuries \citep{hoyt98,vaquero07,svalgaard13}. However, information on both sunspots and sunspot groups, enabling reconstruction of the composite itself is available from 1749 onwards only (for monthly means). 

Continuous daily sunspot information allowing the computation of a daily composite Sunspot Number $S_{\mathrm{N}}$  started in 1848 when R. Wolf initiated systematic sunspot observations in Z\"urich. Between 1749 and 1848, the Sunspot Number was computed from primary standard observers \citep{friedli16}. After 1848, the standard observers were the successive directors of the Z\"urich Observatory (Wolf, Wolfer, \textit{etc}.). Their daily observations were always taken as the reference value, while the best value from secondary stations was chosen when there was no daily observation available from Z\"urich. From 1877 onward, secondary values were an average of all secondary stations instead of the single best value. After 1926, the Z\"urich primary value was an average of the different observers (including assistants) at the main observing station, instead of only the main standard observer.

%%%%%%%%%%% table 1 
\begin{table}
\caption{Key dates in sunspot number observations (1700\,--\,2015). Before 1848, the number of observations per day is highly variable. For that period, information can be found in \citep{hoyt98,vaquero09,svalgaard16,cliver16} \bigskip}
\label{table_dates}
\begin{tabular}{llllc}     
\hline
Period  & Description &  Method  & Time  & Standard  \\
 &  &    & Coverage & Observer \\
\hline
1700\,--\, & Historical sources & Crude yearly & Yearly & - \\ 
1748 & gathered by R. Wolf& averages &  & \\ \hline
1749\,--\,& Historical sources & Standard and & Monthly only  & Staudacher  \\ 
1817& gathered by R. Wolf & auxiliary observer &  &   \\
& & (covering gaps) & & \\ \hline
1818\,--\, & Historical sources,  & Idem & Daily, with  & Schwabe \\ 
1847 & Z\"urich observatory &  & partial coverage  &  \\ \hline
1848\,--\, & Start of systematic  &  Idem,  & Daily, with full & Wolf \\ 
1876 & observations & crude counts  & coverage, one     &  \\ 
     & by R. Wolf   & one to five   & secondary station &  \\ 
     &              & observers     & used for gaps     &  \\ \hline
1877\,--\, & Systematic  & Two standard  & Daily average & Wolf \\ 
1892& observations, & observers   & of secondary & Wolfer \\ 
& full counts & + auxiliary & stations for gaps &  \\ \hline
1893\,--\, & Systematic &  Idem & Daily average & Wolfer \\ 
1927& observations &  & of secondary &  \\
 &  &  & stations for gaps &  \\ \hline
1926\,--\, & Z\"urich network & Several standard & Daily average &  Brunner  \\ 
 1980 	&  	& observers 	& of standard 	  		&  Waldmeier \\ 
 		&  	& + auxiliary 	& observers; a few   	&  \\ 
		&	&				& to $\approx$ 40 stations & \\ \hline					
1981\,--\, & SILSO network & Full network, & Daily average  &  Cortesi \\ 
 2015 &  & pilot station,  & $\approx$ 50 $\rightarrow$  $\approx$ 80 & weighted \\
 &  & outlier removal  & stations &  counts \\ \hline
2015 & SILSO network, & Full network, & Daily average & Cortesi \\ 
 & transition to & pilot station, & $\approx$ 90 stations  & unweighted  \\
 & version 2.0 & outlier removal  & & counts \\
\hline
\end{tabular}

% \makebox[1mm][l]{Before 1848, the number of observations per day is highly variable. For this period, information}\\
% \makebox[1mm][l]{can be found in \citep{hoyt98,vaquero09}}\\
% \makebox[1mm][l]{\citep{svalgaard16,cliver16}}\\
\end{table}

Table~\ref{table_dates} summarises the main changes in time coverage, observers, and observational practices that could have influenced the quality of the observations or the computation of the Sunspot Number. Key dates are taken from \citet{bray64,clette14,friedli16}. 

% \subsection{The sunspot network today}

Between the transfer in 1981, by A. Koeckenlebergh, of the Sunspot Number production from Z\"urich to Brussels, and until 2005, when an extensive reviewing of the routines started, the reduction techniques described by \citet{clette07} did not change significantly. However, the transition to Brussels brought a major change in the production, with S. Cortesi, who is still observing as of July 2016, becoming the standard observer. From 1981 onward, daily values  started to be derived from the whole network instead of a mix of standard observers, with the network as backup. The composite was named ``international Sunspot Number'' ($S_{\mathrm{N}}$) in contrast to the records made by individual observers, called Wolf numbers or relative sunspot numbers\footnote{In the following, ``sunspot number'' will designate the sunspot number in a generic way, whereas ``Wolf number'' will refer to the relative sunspot number $\mathrm{N_W}$ from specific stations, and ``international Sunspot Number'' will be used for the composite $S_{\mathrm{N}}$ from SILSO.}.

The most notable changes occurred in the size and geographical extension of the observing network. The number of observers grew steadily from 1981 to 1995, and then started to decline after Koeckenlebergh's initial phase of intensive recruitment. This number stabilised after 2000  \citep[][Figure 8]{clette14}. From 2005 to 2015, there was an intensive phase of growth in terms of products. In 2015, the international Sunspot Number was finally revised, and new corrections were applied to it; see \citet{clette14,clette15}. Today, the international Sunspot Number is produced and distributed by the Centre for Sunspot Index and Long-term Solar Observations (SILSO, \url{www.sidc.be/silso}), at the Royal Observatory of Belgium.

\subsection{Errors in the Sunspot Number: What we Know}
\label{sec_errors_what}

Although considerable attention has been given to the calculation of the international Sunspot Number \citep{clette15}, its uncertainties remain elusive.  These uncertainties are important not just for statistical assessments of the sunspot number, but also for intercalibrating different Wolf number series and properly merging multiple observations into one single composite record. Few have investigated that question, probably because uncertainty estimation is a non-trivial task. 

There are several aspects to these uncertainties. Most studies concentrate on the effect of random errors, and how close independent measurements are when made under identical conditions. These are usually called precision or repeatability \citep{NIST94}. Accuracy is a different concept, which refers  to differences in absolute calibration, and in instrument bias. Stability is associated with long-term drifts. Both accuracy and stability are taken care of by the scaling factors k; they are beyond the scope of our study, which will mostly concentrate on short-term effects, i.e. the precision. In the following, we shall use the generic word ``error'' to refer to these short-term uncertainties.

\citet{vigouroux94} were among the few who studied the dispersion of daily values within different ranges of monthly sunspot numbers. They concluded that a Poisson-like statistic was appropriate, because the error scaled as the square-root of the sunspot number. Other studies \citep{morfill91} had also reported such a scaling. More recently, \citet{usoskin03} found that all monthly values of the group sunspot number associated with a certain level of daily values also tend to follow a Poisson distribution. 

In an effort to better predict the sunspot number series on time-scales of weeks to years, several statistical models have been developed. \citet{allen10}, for example, used stochastic differential equations to model the sunspot numbers as a diffusion process, whereas \citet{noble11} used a more elaborate Fokker--Planck equation. A direct consequence of such a diffusion process is the exponential distribution of the first-order difference of the sunspot number record \citep{pop11,noble13}. Such methods could potentially be used as well to characterise random fluctuations on a daily basis. However, since they have been primarily designed to reproduce solar variability, and not observational errors, there remains an important need for estimating errors while making few assumptions. 

If the Wolf number $\mathrm{N_W}$ truly behaved like a Poisson variable, with values fluctuating independently from one day to another, then for a fixed level of solar activity this series of daily values should vary randomly with a standard deviation (i.e. error) $\sigma_{N_{\mathrm{W}}} = \sqrt{N_{\mathrm{W}}}$, and we would know the error. In practice, we find a more complex scaling, which can be approximated by $\sigma_{N_{\mathrm{W}}} \approx \alpha \sqrt{N_{\mathrm{W}}}$, with $0 < \alpha \neq 1$.  Most of what follows will concentrate on the estimation of $\sigma_{N_{\mathrm{W}}}$,  its dependence on the solar cycle, and on the observatory. As we shall see, these statistical properties give new insight into the sunspot number.

\subsection{Errors in Sunspot Numbers: Origin}
\label{sec_errors_origin}

Random errors in sunspot numbers may have various origins. The Sun is a natural cause: small spots emerge and wane on time-scales between hours and weeks, while large groups can survive for months \citep{howard92}. Their complex evolution generates a natural scatter in the computed Wolf numbers, with occasional large day-to-day variations.  Because the Wolf number $\mathrm{N_W}$ mixes spots and groups, which have different dynamics, the resulting error becomes a complex function of solar activity. This variability generates natural scatter between non-simultaneous observations.

\begin{figure}[htbp] 
\centerline{\includegraphics[width=0.98\textwidth]{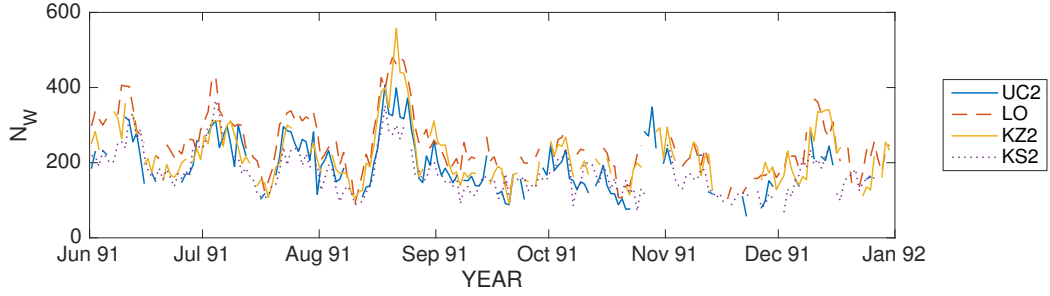}}
\caption{Six months of daily Wolf numbers, as recorded by four stations, illustrating the discrepancies that may arise between observers. The stations are : Uccle Obs. Belgium (UC2), Locarno Obs., Switzerland (LO), Kanzelh\"ohe Obs., Austria (KZ2), and Kislovodsk Obs., Russia (KS2).}
\label{fig_discrepancies}
\end{figure}

A second natural culprit is the observer himself. Two observers who are counting spots and groups simultaneously are likely to disagree because seeing conditions may not be the same, telescopes have different resolutions, their procedure of counting spots and splitting groups differ, and so on. Some of these effects may also result in a weakly nonlinear dependence between the numbers of spots counted by two different observers \citep{lockwood16}. In addition, although most observers report a unique observation per day, some may take the best observation among several. In practice, these discrepancies between observers can reach several tens of percent. Even after correcting them for the scaling factor k, discrepancies remain in the temporal evolution of the daily Wolf number; see Fig.~\ref{fig_discrepancies}.

%%%%%%%%%%%%%%%%%%%%%%%% FIGURE

Disentangling the origins of these errors is a challenging task that goes beyond the scope of our study. Let us therefore generically call error whatever causes the sunspot number to deviate from its true value.

Such errors can be assessed either in the time domain, or by comparing observers. ``Time domain errors'' can be inferred by time series analysis of Wolf number records. This is possible only when these records are sufficiently long and uninterrupted. Differences between observers, called here ``dispersion errors'', on the contrary require a substantial sample of simultaneous observations, with no need for continuity in time. 

Thanks to the large number of stations that offer long and almost uninterrupted records of their Wolf number, we are now in a unique position of inferring and comparing these two types of errors from several decades of observations. To do so, we consider a dataset of 52 quasi-continuous observations, most of which are used for compiling the international Sunspot Number at SILSO. More details on these observations can be found in Table 1 of the review by \citet{clette15}. In the following, we consider daily values from 2 January 1944 to 8 February 2015. For the international Sunspot Number we use the latest version, which is currently version 2.0.

This article is structured as follows: Section~\ref{sec_estimating} presents different estimators of the error, while Section~\ref{sec_comparing} compares their results. In Section~\ref{sec_what} we discuss their implications on the Sunspot Number, and in Section~\ref{sec_transforming} we propose a transform to ease their interpretation. Section~\ref{sec_consequences} focuses on the  consequences of this analysis. Conclusions follow in  Section~\ref{sec_conclusions}.

%%%%%%%%%%%%%%%%%%%%%%%%%%%%%%%%%%%%%%%%%%%%%%%%%%%%%%%%%%%%%%%%%%%%%%%%%%%%%%%%%%%%
% section

\section{Estimating Errors}
\label{sec_estimating}

To assess errors in the different Wolf number series, one should ideally compare the observed sunspot number x(t) to a reference value s(t)
\begin{equation}
x(t)  = \lambda s(t) + \eta(s(t),t) 
\label{eq_error_definition}
\end{equation}
where $\eta$ incorporates both additive and multiplicative errors, and the factor $\lambda$ is normally compounded by the use of the time-dependent scaling factor k, so that $\lambda=1$. Since there is no such reference sunspot number, errors can only be guessed by making assumptions. In the following, we shall generically call $\eta$ the residual noise, and its standard deviation the error.

There are two complementary approaches for estimating $\eta$. One is by comparing simultaneous observations provided by different stations. These so-called ``dispersion errors'' will be addressed in Section~\ref{sec_dispersion_error}. Let us first start with ``time-domain errors'', which consist in determining how regularly a given Wolf number record evolves in time.

%%%%%%%%%%%%%%%%

\subsection{Time-Domain Errors}
\label{sec_time_domain_errors}

In the following, we consider four approaches for estimating time-domain errors.

\subsubsection*{Spectral Noise Floor Model}

When we mix a time series s(t) with a sequence of uncorrelated white noise $\mathrm{\eta(t)}$ of standard deviation $\sigma_{\eta}$, then the power spectral density of the mixture $\mathrm{x(t) = s(t) + \eta(t)}$ equals 
\begin{equation}
P_x(\omega) = P_s(\omega) + P_{\eta}(\omega) = P_s(\omega) + \sigma_{\eta}^2 \ ,
\end{equation}
where $P_i$ stands for the power spectral density (with proper normalisation) of time series $i$. The power spectral density of the sunspot number record tends to fall off rapidly with frequency because of the lack of variability on sub-daily time-scales. As a consequence, additive noise will manifest itself by a floor level at high frequency. Such a noise floor  provides a simple means for quantifying $\sigma_{\eta}$, assuming that the noise is white. For coloured noise, whose power spectral density $\mathrm{P(\omega)}$ varies as $\mathrm{\omega^{-\gamma}}$, this estimator remains valid, but it then applies only to the high-frequency tail of the spectrum. 

Two advantages of this estimator are its simplicity, and the possibility to handle time series with irregular sampling. However, the spectral estimator may underestimate the true noise level, especially if the noise spectrum is non-white (i.e. for $\gamma > 0$). Here we estimate the power spectral density by means of the Lomb--Scargle method \citep{press89}, which does not require regularly sampled records.

\subsubsection*{Autoregressive Sunspot Number Model}

The idea behind autoregressive (AR) modelling is to reproduce the dynamics of the sunspot number time series with a linear parametric model, and then use that model to forecast the sunspot number one day ahead. The innovation, or difference between the observed and the predicted sunspot number then provides us with an estimate of the residual noise $\mathrm{\eta(t)}$. Linear parametric models are widely used to model systems that can be described by linear differential equations with a stochastic forcing \citep{chatfield03}, and AR models have been shown to properly capture coloured and white noise in climate data \citep{mann96,schulz02}.

We model the daily-valued sunspot number time series $\mathrm{x(t)}$ as
\begin{eqnarray}
\hat{x}(t_{i}) &=& a_1 x(t_{i-1}) + a_2 x(t_{i-2}) + \cdots + a_p x(t_{i-p}) \label{eq_AR_1}  \\
\eta(t_{i}) &=& x(t_{i}) - \hat{x}(t_{i}) \ , \label{eq_AR_2}
\end{eqnarray}
where $\mathrm{\hat{x}(t)}$ stands for the predicted sunspot number. We find that models of order $p=8$ yield the best compromise between model parsimony and predictive capacity. High-order models tend to become unstable, and, conversely, the predictive capacity drops for small orders. Note that the time-average of x(t) must be subtracted before fitting the AR model.

This estimator of the noise level requires time series that are regularly sampled, and devoid of data gaps. To meet this requirement, we first interpolate all data gaps by expectation-maximization \citep{ddw11f}. This powerful technique relies on the high correlation between simultaneous observations of the sunspot number to replace missing values by conserving the values of the covariance matrix between observations, regardless of whether there are observations or not. Expectation-maximization has been widely used in climate data analysis \citep[e.g.][]{schneider01}, and in addition it allows the interpolation error to be estimated. Here, we first fill in all data gaps, then estimate the AR model, and subsequently consider the residual error $\mathrm{\eta(t_{i})}$ only for those days $\mathrm{t_{i}}$ for which there is an observation.

The main asset of the AR model is the possibility to estimate the residual noise on short time intervals (typically, a few months), which we shall make use of to investigate solar cycle variations of $\mathrm{\eta(t)}$.

\subsubsection*{Autoregressive Noise Model}

The sunspot number record $\mathrm{x(t) = s(t) + \eta(t)}$ can be viewed as random noise $\mathrm{\eta(t)}$ superimposed on a slowly fluctuating signal component $\mathrm{s(t)}$ with large excursions over decadal periods. In that case, it is appropriate to difference the data to remove that large-amplitude slow component. 

Let $\mathrm{y(t_i) = x(t_i) - x(t_{i-1})}$. For small time steps we have $\mathrm{s(t_i) \approx s(t_{i+1})}$, so that
\begin{equation}
y(t_i) = \eta(t_i) - \eta(t_{i-1}) \ .
\end{equation}
Broadband noise is often well modelled by first-order AR models, in which the predicted value reads
$\mathrm{\hat{\eta}(t_i) = a_1 \eta(t_{i-1})}$. From this we obtain
\begin{eqnarray}
\langle y(t_i) y(t_i) \rangle &=& \; 2\sigma_{\eta}^2(1-a_1)  \\ 
\langle y(t_i) y(t_{i+1}) \rangle &=& -\sigma_{\eta}^2(1-a_1)^2 \nonumber
\end{eqnarray}
where $\langle \cdots \rangle$ stands for ensemble averaging. Finally the standard deviation of the noise becomes
\begin{equation}
\sigma_{\eta} = \frac{\langle y(t_i) y(t_i) \rangle}{2 \sqrt{\langle y(t_i) y(t_{i+1}) \rangle}} \ .
\end{equation}

The requirements and assets of this approach are the same as  for the regular AR model. However, because it is unaffected by trends and slow variations, the differenced AR model is likely to give a more realistic estimate of the noise level.

\subsubsection*{Wavelet Denoising}

Wavelet denoising \citep{ogden96} consists in decomposing a record into discrete series of non-redundant wavelet coefficients that describe the spectral content at different time-scales. Random noise tends to be evenly spread out in time, and over all wavelet coefficients, whereas the salient features of the signal of interest are generally captured by a few outstanding wavelet coefficients, by virtue of the properties of the wavelet transform. Wavelet denoising then consists in thresholding these coefficients: the smallest ones are discarded, while the largest ones are retained for reconstructing the signal. Conversely, we recover the residual error by using the smallest wavelet coefficients only for reconstruction.  

Since we are interested in daily variations only, we consider wavelet coefficients at the lowest level (or time-scale) only. As shown by \citet{donoho93}, a robust measure of the noise level at the smallest scale is then given by 
\begin{equation}
\sigma_{\eta} = 0.675 \; \textrm{median}_i(|c_i^{j=1}|) \ ,
\label{eq_wavelet}
\end{equation}
where $\mathrm{\{ c_i^{j=1} \}}$ denotes the wavelet detail coefficients obtained with a single-level (j=1) discrete-wavelet decomposition. The median is performed over the time indices i, and it attenuates the impact of large outliers.  In the following we consider fourth order Daubechies wavelets, which provide a good compromise between compactness and regularity. The normalisation by 0.675 ensures that for white noise with unit variance, we obtain $\mathrm{\sigma_{\eta} = 1}$.

As for the AR model, our wavelet noise estimator requires daily sampled records with no data gaps. To overcome this, we interpolate missing values as with the AR model, and in Equation~\ref{eq_wavelet} we ignore all wavelet coefficients that are associated with data gaps.

This multiscale approach quantifies the ubiquitous level of random fluctuations that affect daily variations of the sunspot number, and it excludes discontinuities, whatever their origin. For that reason, the wavelet noise estimator is likely to underestimate the noise level.

%%%%%%%%%%%

\subsection{Comparison of Time-Domain Errors}

\begin{figure}[htbp] 
\centerline{\includegraphics[width=0.98\textwidth]{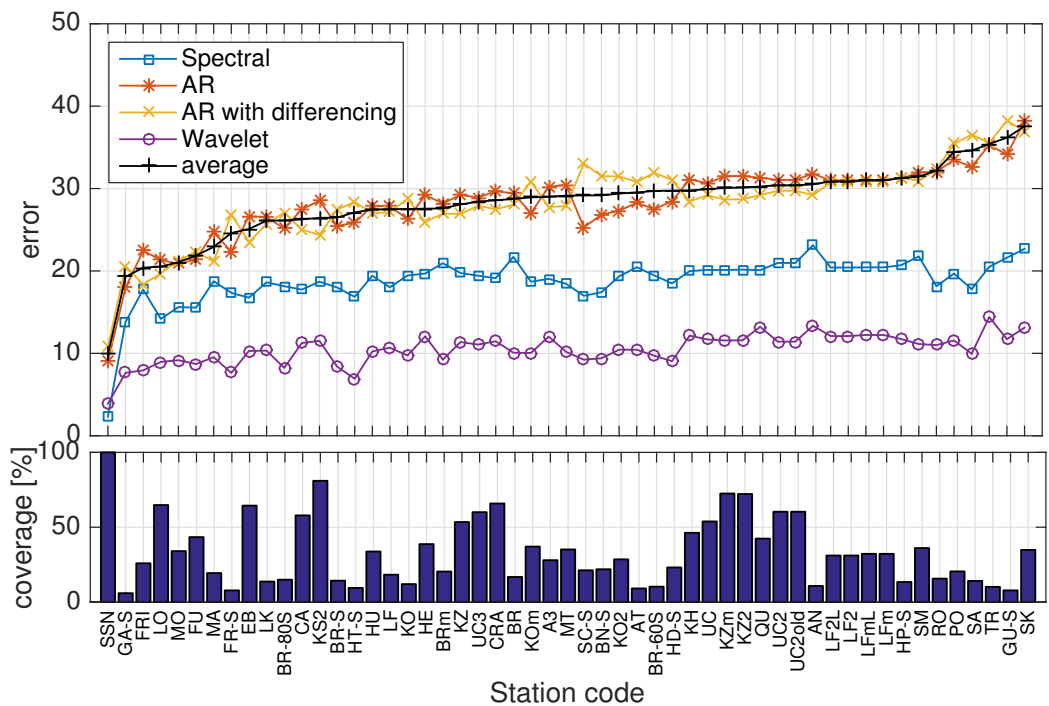}}
\caption{Average error in Wolf number records as estimated by four different methods: i) spectral method, ii) AR method, iii) AR method with differencing, and iv) wavelet method. The  histogram represents the time-coverage of these observing stations. The stations are sorted after the average of the two AR methods. The codes refer to the sample of 52 stations studied by \citet{clette15}; SSN stands for the international Sunspot Number, which is the only one to have 100\, \% coverage.}
\label{fig_errors_comparison}
\end{figure}

Figure~\ref{fig_errors_comparison} compares the time-domain errors in the sunspot number as obtained by the four methods with our sample of 52 stations. As expected, the wavelet estimator, which ignores outliers, yields the smallest errors. The two AR methods, on the contrary, yield the largest estimates. This was also expected, as these methods are better suited for capturing random fluctuations with a non-white power spectral density. The close agreement between the two AR methods suggests that long-term variations do not affect their outcome.

The errors that we obtain by the four methods are within a factor of 2.5 to 3 of each other, which is reasonably close, given their differing assumptions. Our preference goes to the AR methods, which are the least likely to underestimate the error and offer good temporal resolution. In Figure~\ref{fig_errors_comparison}, we use the average of the two AR estimates as a single measure of error, and sort all stations after it in order to better reveal the common trend in all error estimates.

The main conclusion that we draw from Figure~\ref{fig_errors_comparison} is the consistency of these error estimates: large values in one estimator usually also lead to large values in the others. Note that the number of observations has no major influence on the error, as suggested by the poor correlation between the errors and the degree of coverage (i.e. number of days with observations versus the total number of days). The ratio between the largest and the smallest average error remains within a factor of two, thus suggesting that all observatories have quite comparable time-domain errors. The error on the international Sunspot Number (labelled here as SSN) is by far the lowest of all, as would be expected from an average.

%%%%%%%%%%%%%%%%%%%%%%%% FIGURE

%%%%%%%%%%%%

\subsection{Dispersion between observers}
\label{sec_dispersion_error}

\begin{figure}[htbp] 
\centerline{\includegraphics[width=0.6\textwidth]{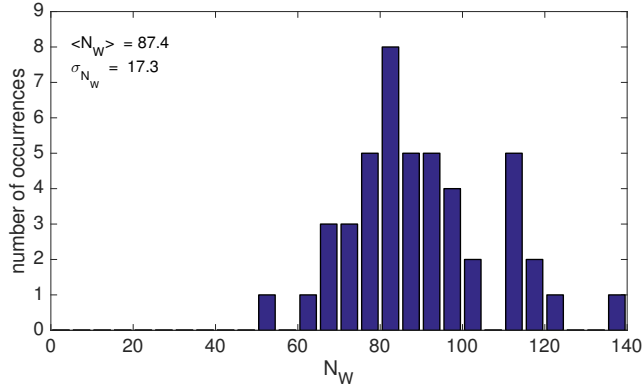}}
\caption{Distribution of Wolf sunspot numbers observed on 1 October 2015 by 46 stations, after correction for their scaling factor k. Most of these values are used for determining the international Sunspot Number.}
\label{fig_histogram}
\end{figure}

This dispersion error requires a  substantial number of concurrent observations, but not necessarily continuity in time.  Figure~\ref{fig_histogram} illustrates it by showing the distribution of sunspot numbers recorded on a given day by 46 stations, after correction by their individual scaling factor k. The international Sunspot Number is based on an average of these values, after eliminating outliers.

%%%%%%%%%%%%%%%%%%%%%%%% FIGURE

\begin{figure}[htbp] 
\centerline{\includegraphics[width=0.98\textwidth]{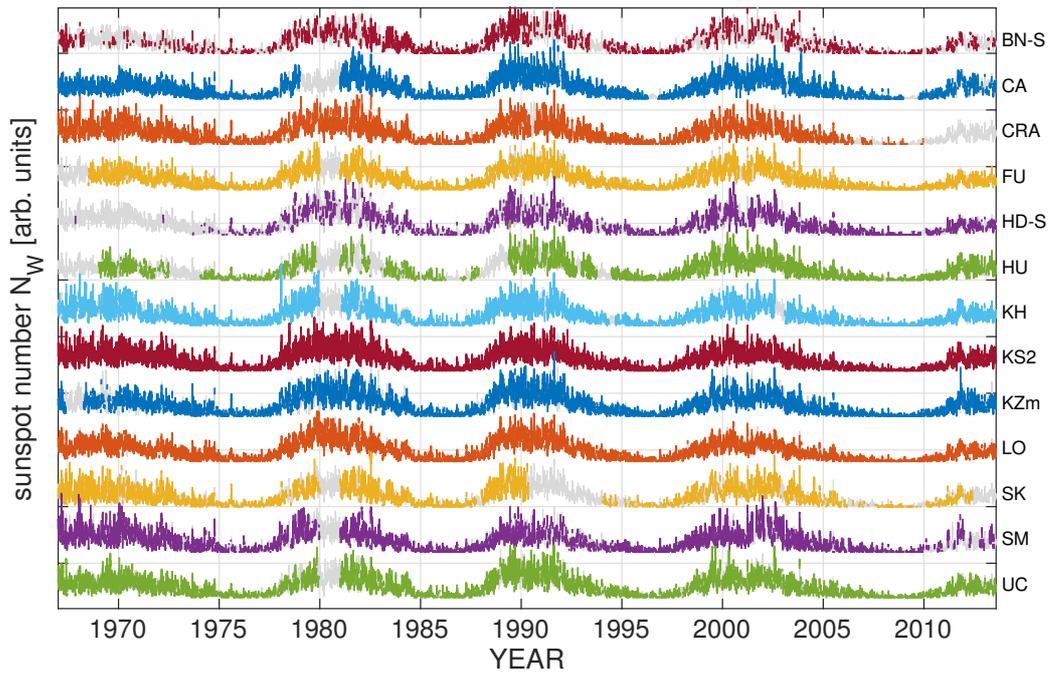}}
\caption{Wolf numbers observed by our subset of 13 stations. Interpolated values appear in grey. The names of the stations, or their individual observers, are: WFS Berlin, Germany (BN-S), Catania Obs., Italy (CA), T.-A. Cragg, Australia (CRA), K. Fujimori. Japan (FU), R. Hedewig, Germany (HD-S), Hurbanovo Obs., Slovakia (HU), Kandilli Obs., Turkey (KH), Kislovodsk Obs., Russia (KS2), Kanzelh\"ohe Obs., Austria (KZm), Locarno Obs., Switzerland (LO), Skalnate Obs., Slovakia (SK), San Miguel Obs., Argentina (SM), Uccle Obs. Belgium (UC).}
\label{fig_selected_stations}
\end{figure}

To properly estimate the dispersion, we select from the sample set of 52 stations a sub-sample of N=13 stations that exhibit high and rather uniform time-coverage between 1967 and 2014. The majority of them have a temporal coverage exceeding 50\, \%.
The Wolf numbers from these 13 stations are shown in Figure~\ref{fig_selected_stations}; the same sample will be used throughout our study. We use the above-mentioned expectation-maximization technique to fill in gaps.

%%%%%%%%%%%%%%%%%%%%%%%% FIGURE

To estimate the dispersion, we first scale all 13 Wolf numbers to the international Sunspot Number $\mathrm{S_N}$ from SILSO:
\begin{equation}
N_{\mathrm{W},i}^*(t) = \gamma_i  N_{\mathrm{W},i}(t) \quad \textrm{where} 
\quad \gamma_i = \frac{S_{\mathrm{N}}}{N_{\mathrm{W},i}}
\end{equation}
because each station uses a slightly different absolute scaling. We estimate the scaling factor $\mathrm{\gamma_i}$ by weighted total least squares \citep{schaffrin08}, and not by classical least squares. The former properly takes into account errors in the numerator and in the denominator. Failing to do so will adversely bias the results; see below in Section~\ref{sec_consequences}.  

The residual error for each station now reads
\begin{equation}
\epsilon_i(t) = N_{\mathrm{W},i}^*(t) - \langle N_{\mathrm{W},j}^*(t) \rangle_j \ ,
\label{eq_dispersion1}
\end{equation}
and its time-dependent standard deviation
\begin{equation}
\sigma(t) = \sqrt{\frac{1}{N-1}\sum_{i=1}^N \epsilon_i^2(t) } 
\end{equation}
provides us with a convenient measure of the dispersion error. We ignore $\sigma(t)$ for days when more than one third of the stations have missing observations. The presence of an excessive number of interpolated values may otherwise lead to an underestimation of high-frequency variations.

%%%%%%%%%%%%%%%%%%%%%%%%%%%%%%%%%%%%%%%%%%%%%%%%%%%%%%%%%%%%%%%%%%%%%%%%%%%%%%%%%%%%
% section

\section{Comparing Time-Domain and Dispersion Errors}
\label{sec_comparing}

\begin{figure}[htbp] 
\centerline{\includegraphics[width=0.8\textwidth]{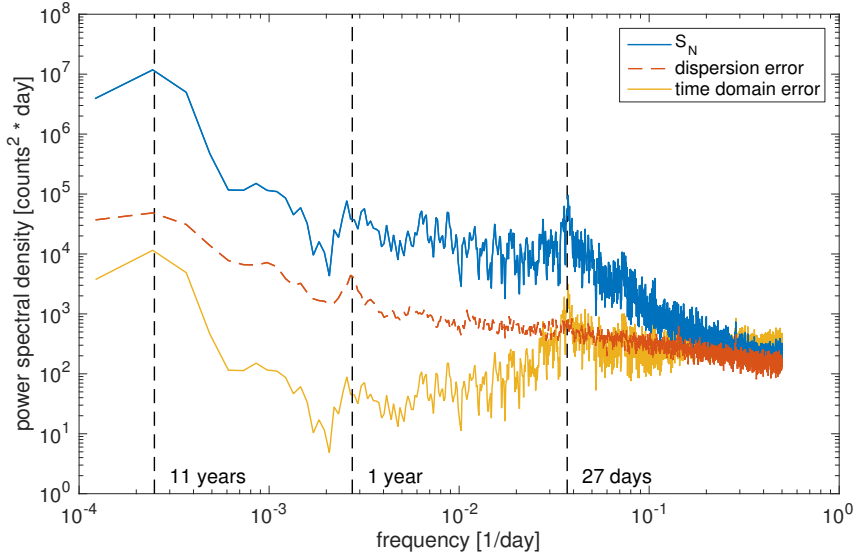}}
\caption{Power spectral density of the sunspot number, of the residual error associated with the dispersion, and of the residual error in the time domain (obtained from the AR model). The latter two power spectral densities are averaged over the 13 selected observers. Three characteristic time-scales are also shown. We estimate the power spectral density by using Welch periodograms.}
\label{fig_psd_residuals}
\end{figure}

Let us briefly compare the properties of the errors before proceeding with their physical interpretation. Figure~\ref{fig_psd_residuals} compares the power spectral density of the residual error in the time domain [$\mathrm{\eta(t)}$ in Equation~\ref{eq_AR_2}], the residual error from the dispersion [$\epsilon(t)$ in Equation~\ref{eq_dispersion1}], and of the international Sunspot Number $\mathrm{S_N(t)}$. The residual error in the time domain mostly contains high frequencies only, as AR models cannot properly capture slow variations in their innovations \citep[see for example][]{ljung97}. The only exception is a conspicuous 11-year modulation that comes from the solar cycle.

%%%%%%%%%%%%%%%%%%%%%%%% FIGURE

The key result here is the close agreement between the two types of errors on short time-scales (typically below 3 to 4 days), which suggests that rapid  variations in are essentially dominated by random noise. For longer time-scales, the signal-to-noise ratio gradually improves, with distinct maxima around periodicities of 27 days and 11 years.

Figure~\ref{fig_psd_residuals} has more to tell. On time-scales shorter than the solar cycle, the power spectral density of the dispersion error on average falls off as a power law, with  $\mathrm{f^{-1/2}}$. Such a scaling suggests that the residual noise is scale-free, behaving similarly to flicker noise. Scale-free behaviour is common in natural processes \citep{sornette04}. However, unlike what has been found in solar studies \citep[e.g.][]{lepreti00}, the scale-free behaviour that we observe deals with the dispersion among observers, and not with the sunspot record itself. Our results suggest that long-range correlations (as caused, for example by slow drifts in the observing strategy) affect the dispersion error. Cases have indeed been reported wherein the sunspot-counting procedure of specific stations has been drifting in time. The use of a scaling factor $\mathrm{k}$ (see Eq.~\ref{eq_Wolf}) is likely to have a major impact on these long-range correlations as it represents a feedback loop in the calibration procedure. The lack of crisp and fully traceable procedure for determining $\mathrm{k}$ is one of the main challenges that awaits the estimation of long-term errors in sunspot number records. 

The dispersion error also reveals some unexpected spectral peaks: the one at a one-year period suggests that the seasonal variations in the observations (the network is mainly located in the northern hemisphere) might introduce a small but statistically noticeable error. Likewise, there is a small peak at seven days, which we might associate with the way data are collected on a weekly basis by some observers. 

Each of the five above-mentioned error estimators (four in the time domain, and one based on the dispersion between observers) comes with its assumptions. In AR models, we assume the residual noise $\mathrm{\eta(t)}$ to be normally distributed. This assumption holds for high levels of solar activity, but it breaks down when the sunspot count drops below approximately 20, where quantisation effects become important. All but the wavelet and dispersion estimators require the noise to be stationary in time, while from the Poisson statistics we expect the error to increase with the sunspot number. These problems can be alleviated by estimating the noise level over shorter time intervals, for which stationarity reasonably applies.

The integer and positive value of the sunspot number bring in additional constraints. The statistical analysis of count rates require a special treatment \citep{davis99}. Ignoring this may affect the outcome of the analysis \citep{bartlett47}. We shall come back to this issue later in Section~\ref{sec_transforming}, and use it there to infer more properties from the sunspot number. Meanwhile, these results tell us that the quantitative comparison of our errors should be done with caution.

%%%%%%%%%%%%%%%%%%%%%%%%%%%%%%%%%%%%%%%%%%%%%%%%%%%%%%%%%%%%%%%%%%%%%%%%%%%%%%%%%%%%
% section

\section{What Errors Tell us about the Sunspot Numbers}
\label{sec_what}

\begin{figure}[htbp] 
\centerline{\includegraphics[width=0.8\textwidth]{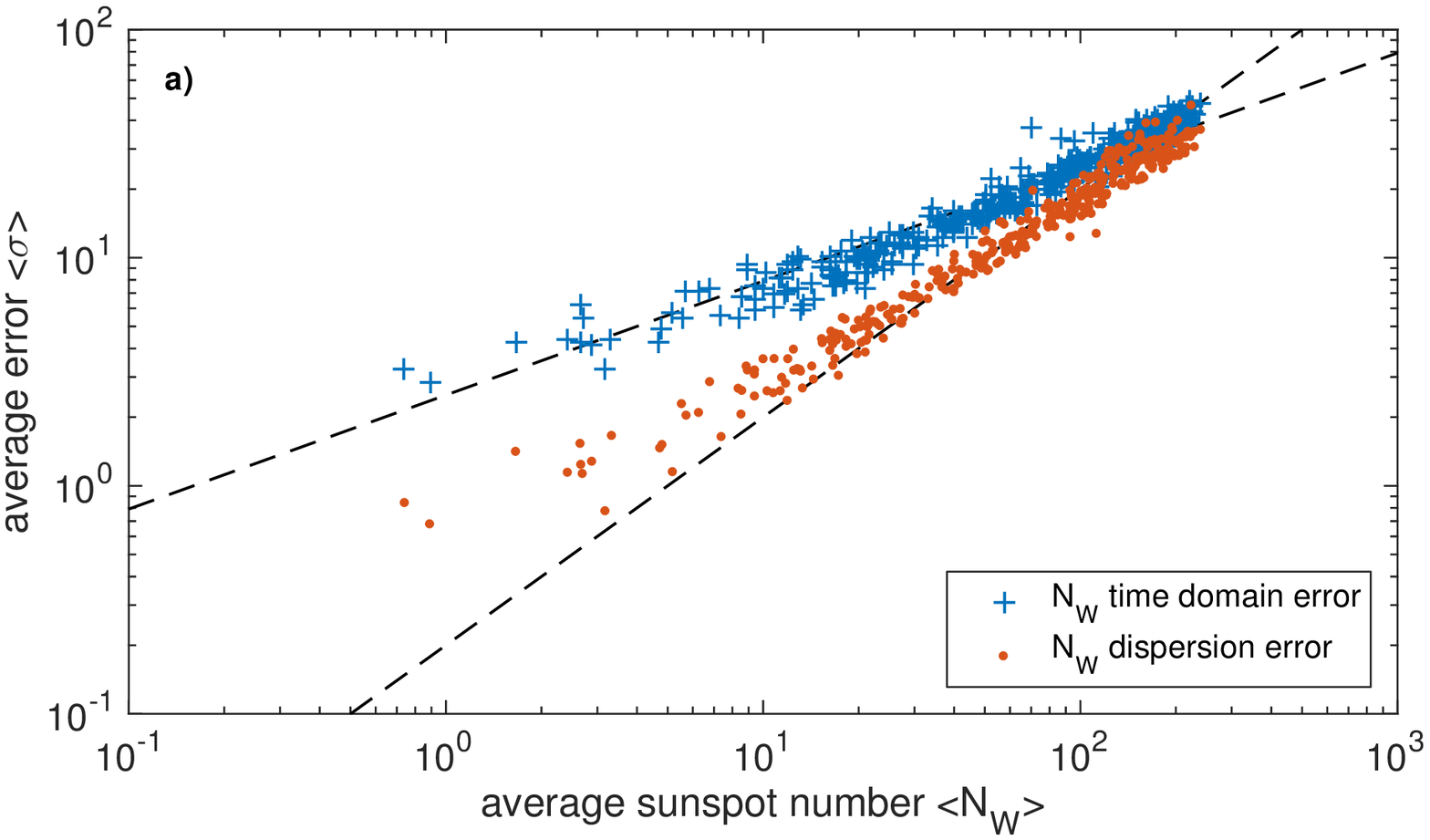}}
\centerline{\includegraphics[width=0.8\textwidth]{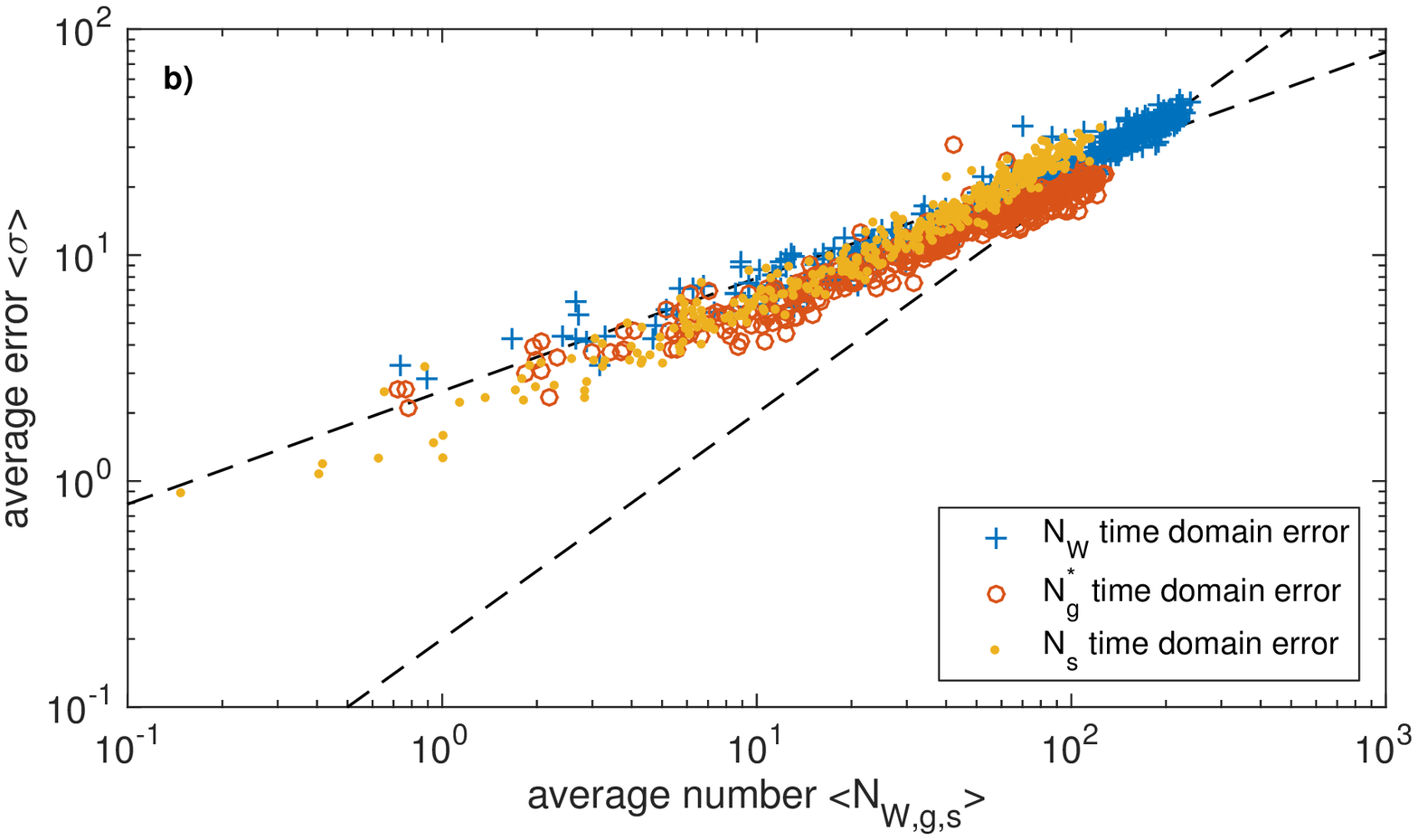}}
\centerline{\includegraphics[width=0.8\textwidth]{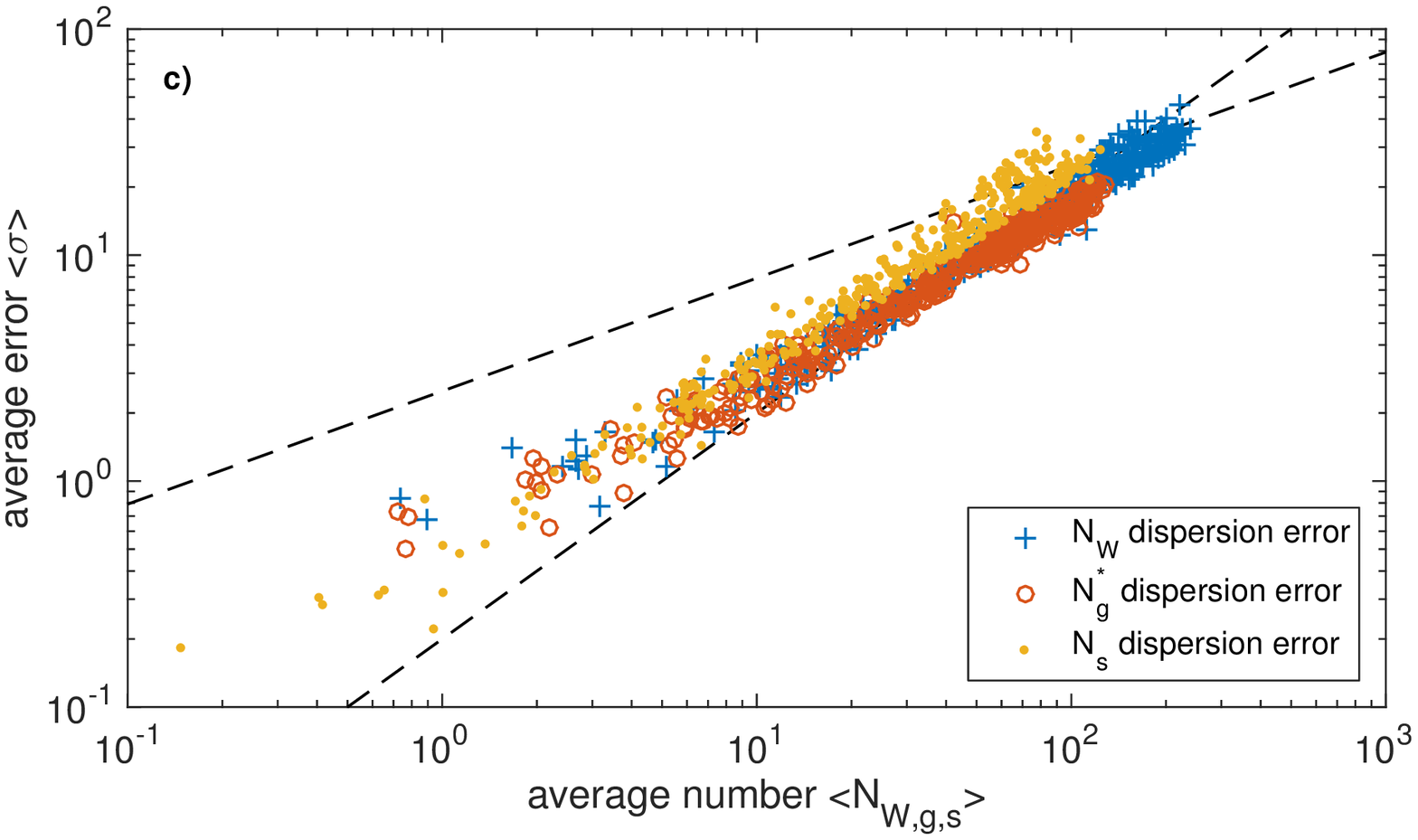}}
\caption{Dependence of the error $\mathrm{\sigma}$ on the Wolf number. Each symbol represents a value averaged over an 82-day time interval. The upper plot (a) compares the time-domain error and the dispersion error of the Wolf number with Wolf number. The middle plot (b) compares the time-domain error of the Wolf number ($N_{\mathrm{W}}$), of the adjusted number of groups ($N_{\mathrm{g}}^* = 10 N_{\mathrm{g}}$), and of the number of spots ($N_{\mathrm{s}}$). The bottom plot (c) shows the same, for dispersion errors. In addition, two scaling laws are shown with dashed lines to facilitate comparison: $\sigma \propto N_{\mathrm{W}}$, and $\sigma \propto \sqrt{N_{\mathrm{W}}}$.}
\label{fig_scaling}
\end{figure}

The central result of this study appears in  Figure~\ref{fig_scaling}, which shows how the error varies with the sunspot number. In this figure, we average the two types of errors and the sunspot number over periods of 82 days; this duration is a compromise between the typical lifetime of an active region and the period beyond which the sunspot number record becomes non-stationary.

%%%%%%%%%%%%%%%%%%%%%%%% FIGURE

Figure~\ref{fig_scaling}a shows how the time-domain error, and the dispersion error scale with the Wolf number $\mathrm{N_W}$. Interestingly, the two errors scale very differently: time-domain errors tend to increase as the square-root of the sunspot number, as one would expect from a Poisson process, whereas the dispersion error increases almost linearly with the sunspot number. A linear scaling would indeed be expected from observation-related errors if these are proportional to the number of spots counted. If, for example, a given observer occasionally tends to underestimate the number of spots by 10\, \%, then this fraction should remain the same, whatever the total number of spots counted. Therefore, a square-root scaling is indicative of solar variability, whereas a linear scaling is more likely to be associated with observational errors.

What seems at first surprising in this figure is the large excess of time-domain errors over the dispersion ones, except near solar maximum. One might expect that the two should be comparable. For sunspot numbers that are typically below 50 to 100, the error is dominated by random fluctuations in time, while most observers tend to agree well and thus have a small dispersion error. We conclude that solar variability, and not the observers, is then the dominant source of the error. This is further corroborated by the scaling of $\sigma$, which then tends to be proportional to the square-root of the sunspot number, for both types of errors. Closer to solar maximum, however, dispersion errors take over, and the scaling switches from a square-root to a steeper linear one. Accordingly, near solar maximum, observational effects, and not the Sun itself, are responsible.

As already highlighted in Section~\ref{sec_errors_origin}, these errors are a complex mix. For example, the ability of observers to count spots is more put to the test during solar maximum, when the distinction of individual spots in large clusters is prone to errors. Whether group counting/splitting is affected in the same way by the solar cycle is unclear; the ambiguity between individual groups and clusters of nearby spots exists whatever the level of solar activity. However, groups are easier to split during solar minimum. To investigate such effects, we focus in Figure~\ref{fig_scaling}b on time-domain errors, but we consider separately the Wolf number ($\mathrm{N_W}$), the number of spots ($\mathrm{N_s}$), and the adjusted number of groups ($\mathrm{N_{\mathrm{g}}^* = 10 N_{\mathrm{g}}}$). In the following, we shall use this adjusted number because it conveniently gives $\mathrm{N_{\mathrm{W}} = N_{\mathrm{s}} + N_{\mathrm{g}}^*}$.

Figure~\ref{fig_scaling}b indeed shows that the error in the number of spots and groups scale similarly, except near solar maximum, where the error in the number of spots tends to grow linearly (and thus faster), while the error in the number of groups continues to grow as the square-root of the sunspot number. We conclude that the number of spots is more prone to observational errors than the number of groups, whose scaling bears the signature of solar variability whatever the level of activity. Accordingly, the number of groups is a statistically better-behaved tracer of solar activity near solar maximum.

Interestingly, the magnitudes of the two types of errors are comparable, which is not a trivial result. Indeed, if the daily number of spots and groups had fluctuated exactly like independent Poisson processes, then we would have $\sigma_{N_{\mathrm{s}}} = \sqrt{N_{\mathrm{s}}}$ and $\sigma_{N_{\mathrm{g}}^*} = \sqrt{10} \sqrt{N_{\mathrm{g}}^*}$. The two errors would then differ by $\sqrt{10} \approx 3$. The difference that we observe between the expected and observed scalings is most likely rooted in the distinct lifetimes of their associated solar structures: groups, on average, have a longer lifetime \citep{howard92}, and therefore are less likely to fluctuate in time, yielding a relatively lower error. 

This different behaviour of groups and spots should also appear in the dispersion error, and Figure~\ref{fig_scaling}c indeed confirms this. According to this figure, observers are more likely to disagree on the number of spots than on the number of groups as soon as the sunspot number exceeds approximately 50. This difference is attenuated when spots and groups are merged into a single sunspot number, although one can still detect a slight increase in the error. 

These results raise several issues and questions. Since $\mathrm{N_s}$ and $\mathrm{N_g}$ have errors that behave statistically differently, one should be extremely careful in regressing one against the other. This issue will be further addressed in Section~\ref{sec_consequences}. We also note that the inflection point observed in Figure~\ref{fig_scaling}a near a sunspot number of about 50 coincides with the transition from simple to more complex spots. \citet{clette15}, when analysing variations in the weighting factors, also observed a modification around that value.

%%%%%%%%%%%%%%%%%%%%%%%%%%%%%%%%%%%%%%%%%%%%%%%%%%%%%%%%%%%%%%%%%%%%%%%%%%%%%%%%%%%%
% section

\section{Transforming the Sunspot Number}
\label{sec_transforming}

One of the main motivations behind our study is to find a simple way to estimate the error at any time and for any station. If the sunspot number truly behaved as a random Poisson random variable, then $\sigma = \gamma \sqrt{N_{\mathrm{W}}}$, with $\gamma=1$ and the problem would be solved. In Figure~\ref{fig_scaling}a, we already see that this scaling does not hold. In addition, $\gamma$ is station- and time-dependent.  

There are several reasons why the sunspot number does not exactly behave like a Poisson variable. First, this would require variations in the number of sunspots to be independent from one day to another, while we know their average lifetime to be considerably longer. Second, the sunspot number mixes groups and spots, and so, a more complex distribution is expected. Furthermore, different observational practices will also inevitably affect the precision.

As a first approximation, one may expect the sunspot number at a given time to be a mix of Poisson and Gaussian random variables, namely $x(t) = \alpha p(t) + g(t)$, with
\begin{itemize}
\item $p \sim \mathcal{P}(\mu_P)$ an independent random Poisson variable whose (time-de\-pen\-dent) expectation is $\mathrm{\mu_P}$. This random variable is multiplied by a gain $\alpha > 0$.
\item $g \sim \mathcal{G}(\mu_G,\sigma_G)$ an independent random Gaussian variable whose expectation is $\mathrm{\mu_G}$, and standard deviation is $\mathrm{\sigma_G}$.
\end{itemize}

The Poisson part $\mathrm{p(t)}$ comes from the counting of fluctuating numbers of sunspots, whereas the Gaussian part $\mathrm{g(t)}$ comes from other additive errors, such as observational ones. $\mathrm{\sigma_G}$ expresses the level of observational noise that gets added to the sunspot number, whereas the gain $\mathrm{\alpha}$ refers to both the lifetime of sunspots and the amount of averaging performed when building the sunspot number record. Smaller values of $\mathrm{\alpha}$ and $\mathrm{\sigma_G}$ are better. 

Most of the statistical tools that we have used so far are optimised for Gaussian random variables, and, therefore, they are not ideally suited for the sunspot number.  This is an incentive for finding a transform that would turn the sunspot number into a new random variable whose distribution is a Gaussian of fixed width and mean value, whatever the level of solar activity. This procedure is called variance stabilisation \citep{bartlett47}. 

There exist several transforms for doing variance stabilisation. The generalised Anscombe transform \citep{anscombe48,makitalo12} is well suited for  a mix of Poisson and Gaussian random variables. The equation 
\begin{equation}\label{eqGAT}
x' = \left\{
  \begin{array}{lr}
    \frac{2}{\alpha} \; \sqrt[]{\alpha x + \frac{3}{8}\alpha^2 + \sigma_G^2 -\alpha \mu_G} & \quad x > -\frac{3}{8}\alpha - \frac{\sigma_G^2}{\alpha} + \mu_G \\
    0 & \quad x \le -\frac{3}{8}\alpha - \frac{\sigma_G^2}{\alpha} + \mu_G
  \end{array}
\right.
\end{equation}
transforms x into a new Gaussian random variable x' whose standard deviation is approximately unity, i.e. $\mathrm{x'} \sim \mathcal{G}(0,1)$. 

If we are able to find $\mathrm{\alpha}$ and $\mathrm{\sigma_G}$, then we have direct access to the error $\mathrm{\sigma}$. Furthermore, $\mathrm{\alpha}$ and $\mathrm{\sigma_G}$ provide insight into the way errors enter the sunspot number.   

The generalised Anscombe transform is frequently used to convert counts into a Gaussian variable that can be more efficiently denoised, before being transformed back. There are more advantages to it. For example, the transformed data are better suited for  AR modelling. Formally, we should apply the AR model after transforming the data. However, this will affect the residual errors, requiring a re-estimation of all quantities. We found the computational price high, for a modest reduction in the residual noise and unchanged conclusions. Therefore, and for the sake of simplicity, we shall not re-estimate the errors here. 

The additive noise that enters the sunspot number has several possible contributions: observer, instrumental, seeing, etc. For that reason, it is unlikely to be biased and we may reasonably set $\mu_G \approx 0$. The generalised Anscombe transform then reduces to
\begin{equation}\label{eqGAT2}
x' =  \frac{2}{\alpha} \; \sqrt[]{\alpha x + \frac{3}{8}\alpha^2 + \sigma_G^2} 
\end{equation}
In the following, we estimate the parameters $\mathrm{\alpha}$ and $\mathrm{\sigma_G}$ by using a sliding window of 11 years to infer the residual error of $x'$ (excluding days with no observations), and then bin this error in intervals of 82 days, exactly as we did in Section~\ref{sec_what}. Finally, we seek the values of $\mathrm{\alpha}$ and $\mathrm{\sigma_G}$ that minimise $|\sigma' - 1|$ for that 11-year interval. The duration of that interval needs to be long enough to let the sunspot number vary from solar minimum to solar maximum, hence the 11-year duration. The gain $\mathrm{\alpha}$ is well constrained by the observations, whereas the value of $\mathrm{\sigma_G}$ is sensitive to what happens near solar minimum, when quantization errors become relatively important; its uncertainty is too large to enable us to discuss it here in meaningful terms.

%%%%%%%%%%%%%%%%%%%%%%%% FIGURE
\begin{figure}[htbp] 
\centerline{\includegraphics[width=0.9\textwidth]{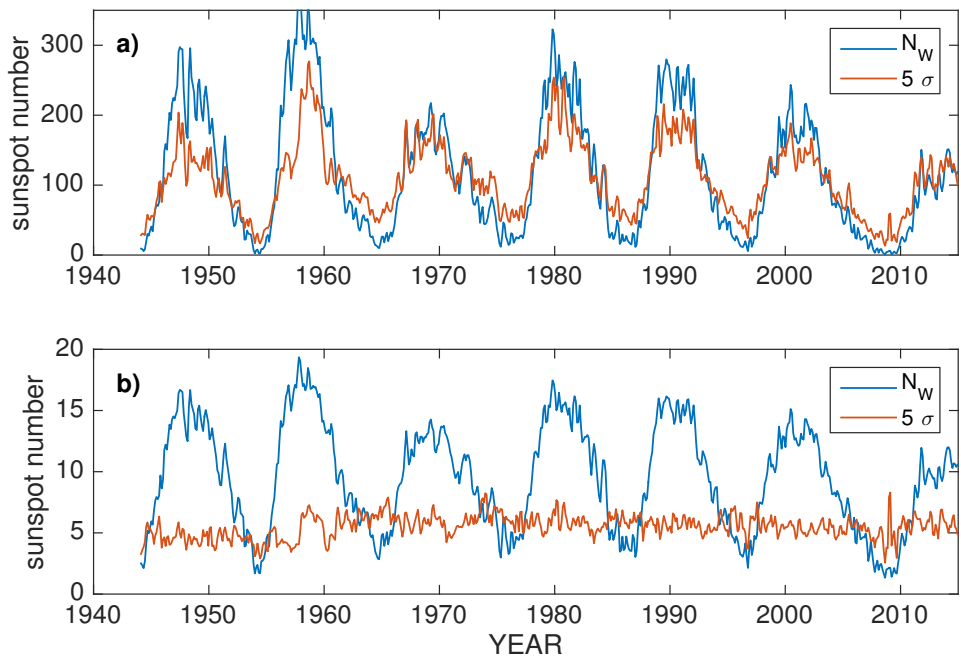}}
\caption{Illustration of the sunspot number and its residual error $\mathrm{\sigma}$ before (a) and after (b) applying the Anscombe transform. The sunspot number record comes from the Locarno station, with $\alpha = 4.2$ and $\sigma_G = 0$. Both the sunspot number and the residual error are averaged over windows of 82 days.}
\label{fig_Anscombe_example}
\end{figure}

Figure~\ref{fig_Anscombe_example} illustrates the usefulness of the generalised Anscombe transform by comparing the sunspot number from the Locarno station, and its time-domain error, before and after the transform. Thanks to this transform, the error now stays remarkably close to unity, regardless of the level of solar activity. This figure \textit{a posteriori} supports the validity of the generalised Anscombe transform.

For the observations from the Locarno station that are shown in Figure~\ref{fig_Anscombe_example}, we find the average gain to be  $\mathrm{\alpha=4.2 \pm 0.2}$ for the Wolf number, $\mathrm{\alpha=2.8 \pm 0.1}$ for the number of groups, and $\mathrm{\alpha=2.9 \pm 0.1}$ for the number of spots. From these, the error can be directly estimated by error propagation; see Equation~\ref{eq_error} below.

%%%%%%%%%%%%%%%%%%%%%%%% FIGURE
\begin{figure}[htbp] 
\centerline{\includegraphics[width=0.5\textwidth]{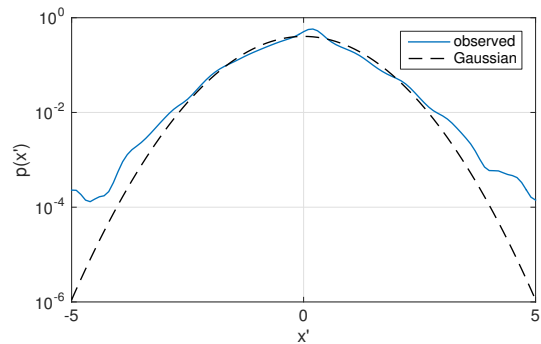}}
\caption{Probability density function of the residual noise of the generalised Anscombe transform of the sunspot number displayed in Figure~\ref{fig_Anscombe_example}. Also shown is a Gaussian distribution of zero mean, and unit variance. Gaussian kernels were used to estimated the probability density function.}
\label{fig_kernel}
\end{figure}

Figure~\ref{fig_kernel} displays the probability density function of the residual noise associated with the transformed sunspot number, confirming that it is indeed close to a Gaussian distribution of unit variance, whatever the level of solar activity. In contrast, the probability density function (not shown) associated with the original data has long tails and is skewed, even when considered for a fixed level of solar activity. The generalised Anscombe transform may be considered here as a renormalisation technique because all residual-noise distributions collapse onto one single curve.

Note that the gain $\mathrm{\alpha}$ is larger for the Wolf number $\mathrm{N_W}$ than for its constituents. Had these been independent, then all values of the gain would have been identical. This difference can thus be ascribed to correlations between $\mathrm{N_s}$ and $\mathrm{N_g}$.  

The gain $\mathrm{\alpha}$ is station-dependent: for the 13 stations that were used in Section~\ref{sec_dispersion_error},  the average gain for the sunspot number ranges from from 4.0 to 7.8. As mentioned before, lower values are indicative of a lower noise level, or may also result from averaging.

Because the error depends on the sunspot number, solar-cycle averages should be treated with care. In Section~\ref{sec_comparing}, we averaged errors over several decades. Such averages are actually difficult to interpret in absolute terms, although they still make sense when comparing methods, or stations, provided that the time span is the same. \\

\begin{figure}[htbp] 
\centerline{\includegraphics[width=0.98\textwidth]{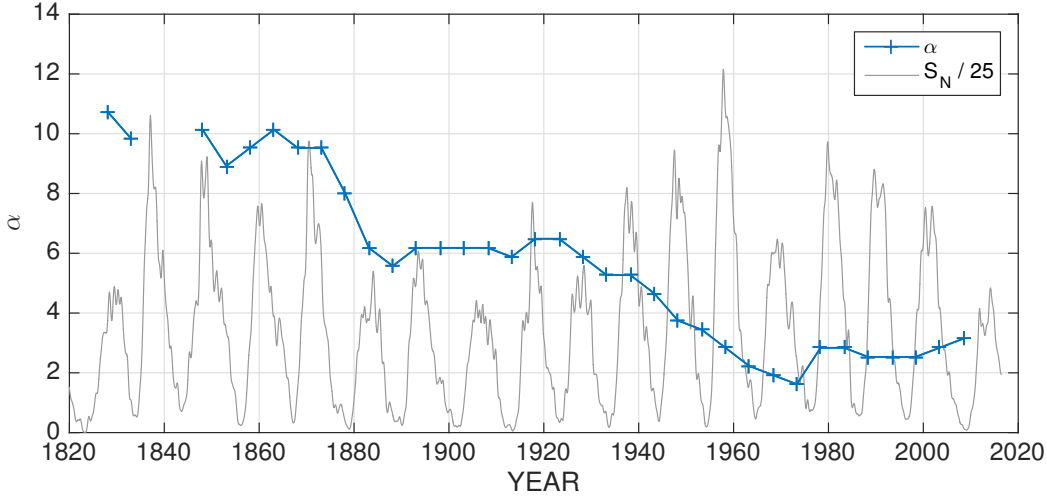}}
\caption{Variation in the gain $\mathrm{\alpha}$, inferred from the international Sunspot Number record, for a sliding window of 11 years (in time steps of 5.5 years).  Missing values occur when the number of data gaps (which cannot be interpolated by other means) exceeds 20\, \%. Also shown is the international Sunspot Number, $S_{\mathrm{N}}$ averaged over six months.}
\label{fig_Anscombe_SN}
\end{figure}

An interesting exercise now consists in monitoring variations in the gain for the international Sunspot Number from SILSO. This sunspot number is a composite record, whose number (and quality) of inputs is time-dependent, and thus it should lead to significant variations in the noise level. Figure~\ref{fig_Anscombe_SN} illustrates this by showing how the gain has evolved from 1818 to 2015. Higher values imply a larger noise level. The sharp drop near 1880 coincides with a change of strategy, when after 1877 the number of sunspots was averaged over two standard observers (Wolf and Wolfer) rather than based on a single daily value. Around 1930, the smaller drop corresponds to the new standard observer Waldmeier. The gradual  drop occurring between 1926 and 1981 is likely due to a combination of factors: i) assistant's values were increasingly often taken into account; ii) values from the network were progressively used to assess the quality of the standard observation; and iii), the network grew during that time period. The lowest gains occur in the 1980s. The transition from Z\"urich to Brussels in 1981 probably led to a slight deterioration in the gain. After 1981, the slight U-shape is concomitant with the change in number of stations/observers. However,  the gain remains stable during that period.   

Figure~\ref{fig_Anscombe_SN} thus vividly illustrates how the handling of multiple simultaneous observations has impacted the quality of the sunspot number. In this sense, it highlights again the need for a statistical assessment of the international Sunspot Number. 

Taking the square-root of the sunspot number is not just a mathematical trick to ease its analysis. In some specific cases, there is also a physical motivation for performing such a transform.  \citet{wang05a} showed that the fluctuations of the solar equatorial magnetic field at 1 AU scale better with the square-root of the sunspot number. More recently, \citet{froehlich16} used the same transform to improve the correlation with the total solar irradiance, and to infer errors from the latter.

%%%%%%%%%%%%%%%%%%%%%%%% FIGURE

%%%%%%%%%%%%%%%%%%%%%%%%%%%%%%%%%%%%%%%%%%%%%%%%%%%%%%%%%%%%%%%%%%%%%%%%%%%%%%%%%%%%
% section

\section{Consequences}
\label{sec_consequences}

Our results have several immediate consequences. First, we are now able to estimate the error in any daily sunspot record, provided that its temporal coverage is sufficiently high, typically $> 70\, \%$. If there are more data gaps, then these need to be filled in beforehand, e.g. by expectation-maximization. By using one of the two AR models (see Section~\ref{sec_time_domain_errors}) we then estimate the residual noise, whose standard deviation provides a measure of the error $\mathrm{\sigma}$.

Alternatively, we may use the generalised Anscombe transform, and, by error propagation, find the error thanks to the relation
\begin{equation}
\label{eq_error}
\sigma =  \alpha \sqrt{\frac{N_{\mathrm{W}}}{\alpha} + \frac{3}{8}} \ .
\end{equation}
Equation~\ref{eq_error} is convenient to use once the gain $\mathrm{\alpha}$ is known. Some values of the gain are tabulated in Table~\ref{tbl_errors} for the international Sunspot Number. From these, and for Cycles 21 and following, we may use the approximation
\begin{equation}
\label{eq_error2}
\sigma \approx  1.7 \sqrt{S_{\mathrm{N}} + 1} \ .
\end{equation}

\begin{table}
\caption{Average gain $\mathrm{\alpha}$ associated with each solar cycle of the international Sunspot Number. Each period is also associated with a standard observer. \medskip}
\label{tbl_errors}
\begin{tabular}{cccc}     
\hline
Solar cycle number & Start and end year & Standard observer &$\mathrm{\alpha}$   \\ \hline
6  & 1810\,--\,1823 & 			&8.9 	\\
\hline
7  & 1823\,--\,1834 &Schwabe 		&10.4\\
8  & 1834\,--\,1844 &				&10.2\\
9  & 1843\,--\,1856 &				&10.2\\
\hline
10 & 1856\,--\,1867 &Wolf 		&9.9 	\\
11 & 1867\,--\,1879 &				&9.3 \\
\hline
12 & 1879\,--\,1890 &Wolf -- Wolfer&6.5  	\\
13 & 1890\,--\,1902 &Wolfer 		&6.3 	\\
14 & 1902\,--\,1914 &				&6.1 \\
15 & 1914\,--\,1924 &				&6.3 \\
16 & 1924\,--\,1934 &				&6.1 \\
\hline
17 & 1934\,--\,1944 &runner 		&5.0 	\\
\hline
18 & 1944\,--\,1954 &Waldmeier 	&3.9 	\\
19 & 1954\,--\,1965 &Cortesi -- 1957&2.9  	\\
20 & 1965\,--\,1976 &				&2.1 \\
\hline
21 & 1976\,--\,1987 &SILSO 		&2.7 	\\
22 & 1987\,--\,1997 &Cortesi		&2.6 	\\
23 & 1997\,--\,2009 &				&2.7 \\
24 & 2009\,--\,     &				&3.1 \\
\hline
\end{tabular}
\end{table}

This error is crucial for making a composite record out of multiple observations. For example, the maximum-likelihood estimator of an average sunspot number reads
\begin{equation}
\label{eq_likelihood}
\langle N_{\mathrm{W}}(t) \rangle = \frac{\sum_{i=1}^N N_{\mathrm{W},i}^{(t)} \sigma_i^{-2}(t)}
{\sum_{i=1}^N \sigma_i^{-2}(t)} \ .
\end{equation}
Given that the error $\mathrm{\sigma}$ typically varies by a factor of two to three between different observers (see Figure~\ref{fig_errors_comparison}), we may expect changes in the international Sunspot Number to occur once these errors are taken into account.

However, a more subtle, and misleading consequence of the errors arises when comparing records. The estimation of the scaling factors k, and the extension of the sunspot number record backward in time rely heavily on comparisons between individual records.  Let us assume that we want to estimate the ratio between the Wolf numbers from the Kanzelh\"ohe (KZm) and Locarno (LO) stations, to be called henceforth $\beta$.

%%%%%%%%%%%%%%%%%%%%%%%% FIGURE
\begin{figure}[htbp] 
\centerline{\includegraphics[width=0.7\textwidth]{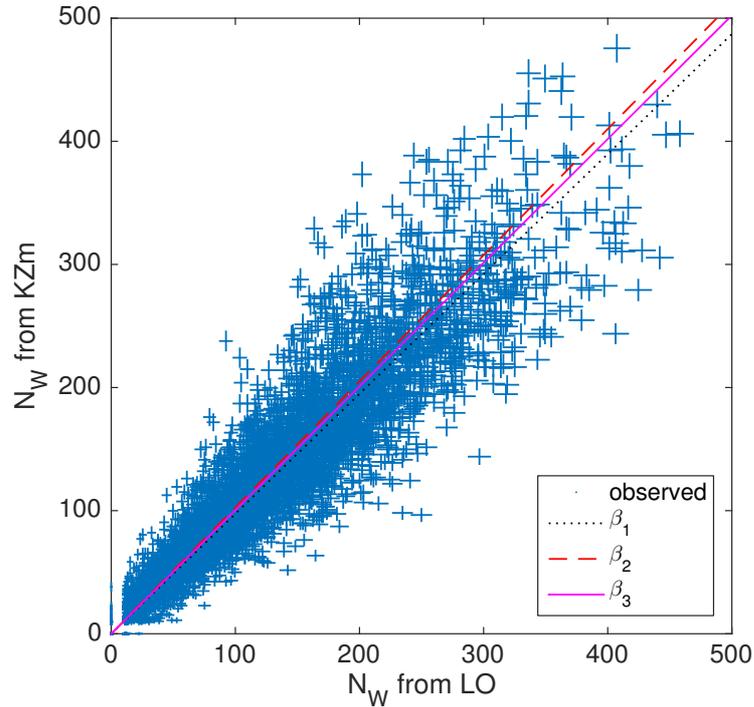}}
\caption{Scatter plot of the sunspot number from the Kanzelh\"ohe Observatory (KZm) \textit{vs} that of the Locarno Observatory (LO). The three fits correspond to the three models, as discussed in the text. Error bars represent $\pm 1 \sigma$.}
\label{fig_regression}
\end{figure}

Figure~\ref{fig_regression} illustrates the scatter plot of these two Wolf numbers and reveals a linear relationship, with considerable dispersion. The ratio between the two numbers can be estimated in different ways. The classical one involves a least-squares fit, wherein we minimise the cost function 
\begin{equation}
J = \sum_i \Big( N_{\mathrm{W,KZm}}(t_i) - \beta_1 N_{\mathrm{W,LO}}(t_i) \Big)^2 \ .
\end{equation}
In doing so, we discard all errors, and assume that the regressor $\mathrm{N_{W,LO}}$ is perfectly known, which is obviously incorrect. 

Alternatively, we could also minimise
\begin{equation}
J = \sum_i \Big( N_{\mathrm{W,KZm}}(t_i)/\beta_2 - N_{\mathrm{W,LO}}(t_i) \Big)^2 \ ,
\end{equation}
wherein we now assume that the dependent variable $\mathrm{N_{W,KZm}}$ is error\-less. Weigh\-ted total least squares (also known as errors-in-variables-regression) is a generalisation of classical least squares, which allows both variables to have errors \citep{golub80,schaffrin08}. Errors-in-variables-regression avoids regression dilution. With it, we obtain a sounder estimate  $\beta_3$ of the ratio, which gives more weight to smaller values.

The ratios we find for the three approaches are respectively $\beta_1 = 0.9729 \pm 0.0068$,  $\beta_2 = 1.0232 \pm 0.0073$, and $\beta_3 = 1.0030 \pm 0.0073$. Notice how the two classical solutions obtained by least squares differ by more than 5\, \%. More importantly, if we regress one record to the other, and back again, then the end result will differ from the initial one, and it will be systematically smaller. Indeed, we have $\beta_1 \beta_2 = 0.9955 < 1$. This is known as the regression-toward-the-mean problem. Small as this difference may be, it will generate a trend in the composite. Present sunspot composites are built mainly with backbones, or by daisy-chaining. With backbones, one or a few stations are preferentially used as references to calibrate all the others \citep[e.g.][]{svalgaard16}, whereas in daisy-chaining each record is regressed to the subsequent one, and calibration thus passes from one station to the other. \citet{lockwood16b} recently highlighted how easily both approaches can generate spurious trends. A natural way out consists in stitching together the different records after decomposing them into different time-scales, while taking into account their uncertainty \citep{ddw16c}.

%%%%%%%%%%%%%%%%%%%%%%%%%%%%%%%%%%%%%%%%%%%%%%%%%%%%%%%%%%%%%%%%%%%%%%%%%%%%%%%%%%%%
% section

\section{Conclusions}
\label{sec_conclusions}

In this study we have provided the first thorough estimate of the uncertainty associated with daily values of the sunspot number. Our main findings are:
\begin{itemize}

\item Estimating the uncertainty (what we call here error) without an independent reference requires assumptions. Among existing techniques, autoregressive models provide a simple, and yet reliable, method for estimating the error based on the difference between the observed and predicted sunspot number. This error mostly quantifies short-term variability, i.e. precision. 

\item We consider two types of errors: a ``time-domain'' one, estimated via autoregressive modelling, and a ``dispersion error'', obtained from the scatter among observers for a given day. The latter is systematically lower for sunspot numbers up to about 100. The two  scale differently with the sunspot number: a square-root dependence (i.e. Poisson-like) is found for the ``time-domain error'', and a more linear one for the ``dispersion error''. We conclude that random fluctuations in solar activity are the prime cause for errors for sunspot numbers up to about 100.  Observational errors prevail beyond that value.

\item The relative error is smaller for the number groups than for the number of spots. In addition, the former has a dominant contribution coming from solar variability, whatever the level of solar activity. In this sense, the number of groups is less likely to be affected by observer effects, and is a more robust quantity.

\item Interestingly, dispersion errors show evidence of scale-free variations, which suggests that differences between observers are affected by long-range correlations, i.e. slow drifts.

\item The generalised Anscombe transform gives us an analytical model for explaining how variations can be described in terms of a mix between Poisson and Gaussian random fluctuations. The former is by far the dominant one, and observational constraints are too weak to properly assess the (presumably weak) contribution from Gaussian fluctuations. Such a transform offers new perspectives for describing solar fluctuations as a diffusion process, as the transformed sunspot number behaves as a Gaussian random variable.

\item For the international Sunspot Number, and for solar cycles after 1981, the error (i.e. precision) can be approximated by $\sigma \approx 1.7 \sqrt{S_{\mathrm{N}} + 1}$, where  $S_{\mathrm{N}}$ is the sunspot number. The two numerical coefficients are time- and observer-dependent. By monitoring their evolution in time, we witness how changes in the observation strategy have affected the error in the sunspot number since the early 19th century.

\item The nonlinear scaling of the error with sunspot number means that linear regressions between different sunspot records should be done with the utmost care. Classical least square is not appropriate, as it tends to bias the results. We recommend instead error-in-variables-regression by total least squares.
\end{itemize}

We have not addressed so far the estimation of errors from monthly, or from yearly sunspot numbers. Indeed, these are quite different issues. In principle, such errors can be simply propagated from daily observations (when available) provided we know their covariance matrix. Unfortunately, the correlation between successive values of the sunspot numbers depends on the different lifetimes of spots and groups, and that of observational errors. In addition, it depends on the level of solar activity. For these reasons, the derivation of errors for monthly or yearly values is a task that requires a separate study.

Before the 19th century, and for periods when the time coverage is sparser, error propagation may not be adequate, and other approaches need to be considered, such as the one described by \citet{usoskin03}, after revisiting some of their assumptions in the light of what we found here. The analysis of such sparse data is likely to become an important issue as more historical records are being uncovered \citep{arlt08,vaquero09}. For such irregular observations, the spectral-noise estimate (see Section~\ref{sec_estimating}) could be a fallback solution. However, this estimator becomes unreliable as the sample size shrinks. 

On time-scales of months and beyond, linear autoregressive models are no longer adequate either. The reason for this is not the sample size, but the requirement to have nonlinear models in order to properly describe the sunspot number on those time-scales \citep[e.g.][]{letellier06}.

Another major challenge is the assessment of the stability of the sunspot number, which is crucial for properly reconstructing past solar activity. Our time-domain estimates are of no help here, because they only detect short-term variations. Our dispersion error diagnoses the presence of long-term errors in the sunspot number, but it cannot remove them. Clearly, this will become an important issue for future revisions of the international Sunspot Number.

%%%%%%%%%%%%%%%%%%%%%%%%%%%%%%%%%%%%%%%%%%%%%%%%%%%%%%%%%%%%%%%%%%%%%%%%%%%
%% Acknowledgements
%

\subsection*{Acknowledgements}
We gratefully acknowledge funding from the European Community's Seventh Framework Programme (FP7, 2012) under grant agreement no 313188 (SOLID, \\ \url{projects.pmodwrc.ch/solid}). We sincerely thank the anonymous referee for carefully reviewing the manuscript, and for making pertinent comments.

%%% %%%%%%%%%%%%%%%%%%%%%%%%%%%%%%%%%%%%%%%%%%%%%%%%%%%%%%%%%%%
%% Bibliography
%
% Using BibTeX
%
\bibliographystyle{plainnat}

%\bibliography{../../../mybib}  
%
% Without BibTeX 
% \begin{thebibliography}{}
% \bibitem[\protect\citeauthoryear{Author}{Year}]{key}
%   <bibliographical entry>
%
% \bibitem[\protect\citeauthoryear{}{}]{}
%   
%  
% \end{thebibliography}

\end{document}